\documentclass[fleqn,twoside]{article}
\usepackage[headings]{espcrc2}
\usepackage{graphicx}
\usepackage{rotating}
\usepackage{amsmath,amssymb}
\usepackage{cite}
\newcommand{\No}{No.}

\title{Energy spectrum and mass composition of primary cosmic rays around
the `knee' in the framework of the model with two types of
sources}

\author{A.\,A. Lagutin\address[ASU]{Theoretical Physics Department, Altai State University,\\
Lenin Avenue 61, Barnaul 656049, Russia}%
\thanks{Work was supported by RFBR grants \No\,04--02--16724, \No\,06--02--27113.},
        A.\,G. Tyumentsev\addressmark,
        A.\,V. Yushkov\addressmark}

\runtitle{Energy spectrum and mass composition}

\runauthor{A.\,A. Lagutin, A.\,G. Tyumentsev, A.\,V. Yushkov}

\begin{document}

\begin{abstract}
Analysis of the experimental data on cosmic ray spectra in the framework of the 
proposed model with two types of sources leads to conclusion, that sources with 
particle generation spectral exponent $p\sim 2.85$ give the major contribution 
to the all-particle spectrum in the energy range $10^5-10^7$~GeV. `Fine 
structure' of spectrum around the `knee' may arise due to presence of nearby 
supernova type source, accelerating particles up to the energies $\sim3\cdot 
10^4 Z$~GeV, if the energy output of such source is $\sim2\cdot10^{48}$ 
erg/source. 
\end{abstract}

\maketitle

\section{Introduction}

Behaviour of primary cosmic ray nuclei spectra in the vicinity of the `knee'
is, undoubtedly, one of the key elements for solution of problem of high energy
cosmic rays sources and for establishment of acceleration mechanisms. So, for
example, non-monotonous behaviour of mass composition and additional breaks in
all-particle spectrum in the energy range $10^5-10^7$~GeV would support
scenario with supernovae as the major cosmic ray sources and acceleration on
shock-waves fronts up to the energies $E_{\max}\sim 10^5Z$~GeV. Particle
spectrum produced by this mechanism may be represented as $S_{\text{SN}}\sim
E^{-2}\Theta(E_{\max}- E)$, where Heaviside step function $\Theta(x)$
qualitatively reflects a sharp cut-off of the spectrum for $E >
E_{\max}$~\cite{Berezh1988,Berezh1996a}.

The recent results on the proton spectrum of Tibet collaboration~\cite{tibet} 
with no indications on cut-off in the spectrum up to $\sim10^7$~GeV allow to 
affirm, that proton spectrum in the considered energy range is formed by 
sources with generation spectrum being different from $S_{\text{SN}}$. On the 
other hand, in papers~\cite{Erlyki2001ic,Shaulo1999} it is shown, that 
all-particle spectrum in this energy range has several breaks. 

Reconciliation of experimental results~\cite{tibet} and 
deductions~\cite{Erlyki2001ic}, based on the analysis of EAS size spectra and 
data on Cherenkov radiation of showers, is possible, in our opinion, only 
within the framework of the following scenario: 1) particle generation spectrum 
for energies $10^5-10^7$~GeV in the principal cosmic ray sources is different 
from that of $S_{\text{SN}}$; 2) contribution of young nearby supernova 
determines `fine structure' of the spectrum and non-monotonous character of 
mass composition behavior in the `knee' region. 

In this paper we present new calculations of energy spectra and mass 
composition of cosmic rays, performed for the above formulated scenario of 
particles generation in two distinct types of sources under assumption of 
anomalous diffusion of particles in inhomogeneous galactic medium. The main 
purpose of the calculations is to determine conditions under which the proposed 
model (two types of sources with different acceleration mechanisms + anomalous 
diffusion) reproduces results of~\cite{Erlyki2001ic,Hoeran2003}.

\begin{figure}
\centering\includegraphics[angle=-90,width=.5\textwidth]{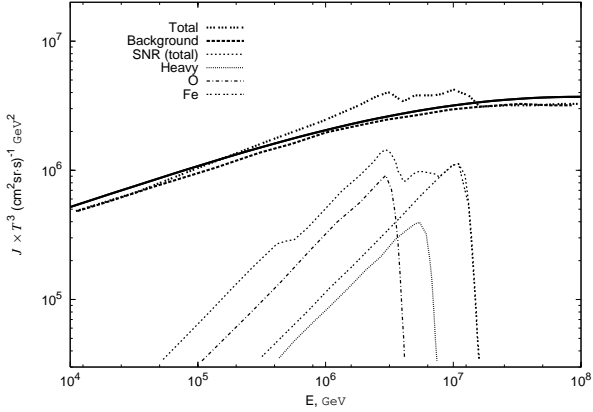}
\caption {Cosmic ray spectrum in the `knee' region. Dotted lines ---
data from~\cite{Erlyki2001ic}, solid line --- calculations in the anomalous
diffusion model with one type of sources $S\sim E^{-2.85}$.}
\label{ew1}
\end{figure}

\section{Diffusion model}

Following to~\cite{Lagutin2003,Lagutin2003an}, cosmic ray
transport in fractal interstellar galactic medium with `traps' and
without energy losses and nuclear interactions is given by
anomalous diffusion equation
\begin{eqnarray}\label{eq:Green}
\nonumber\frac{\partial N}{\partial
t}=-D(R,\alpha,\beta)\mbox{D}_{0+}^{1-\beta}(-\Delta)^{\alpha/2}
N(\vec r,t,R)+\\ +S(\vec r,t,R).
\end{eqnarray}
Here $D=D_0R^\delta$ -- anomalous diffusion coefficient, $R$ --- particle 
rigidity, $\mbox{D}_{0+}^\mu$~---~the~Riemann-Liouville fractional 
derivative~\cite{Samko1987}: 
\begin{equation*}
\mbox{D}_{0+}^\mu f(t)\equiv
\frac{1}{\Gamma(1-\mu)}\frac{d}{dt}\int\limits_0^t
(t-\tau)^{-\mu}f(\tau)d\tau,\ \mu<1,
\end{equation*}
$\left( -\Delta \right)^{\alpha/2}$~---~fractional Laplacian (`Riss' 
operator)~\cite{Samko1987}: 
\begin{equation*}
(-\triangle)^{\alpha/2}f(x)=\frac{1}{d_{m,l}(\alpha)} 
\int\limits_{R^m}\frac{\triangle^l_y f(x)}{|y|^{m+\alpha}} dy,
\end{equation*}
where $l > \alpha$, $x \in {\rm R}^m$, $y \in {\rm R}^m$, 
\begin{equation*}
  \Delta_{y}^{l} f(x) = \sum_{k=0}^{l} (-1)^{k} {l\choose
k} f(x-ky) 
\end{equation*}
\begin{equation*}
 d_{m,l}(\alpha) =
\int\limits_{{\rm R}^m}(1-e^{iy})^l |y|^{-m-\alpha}dy.
\end{equation*}
If $\alpha=2$ and $\beta=1$, the equation~(1) is just the normal diffusion 
equation. 

\begin{figure*}
\centering\includegraphics[angle=-90,width=.7\textwidth]{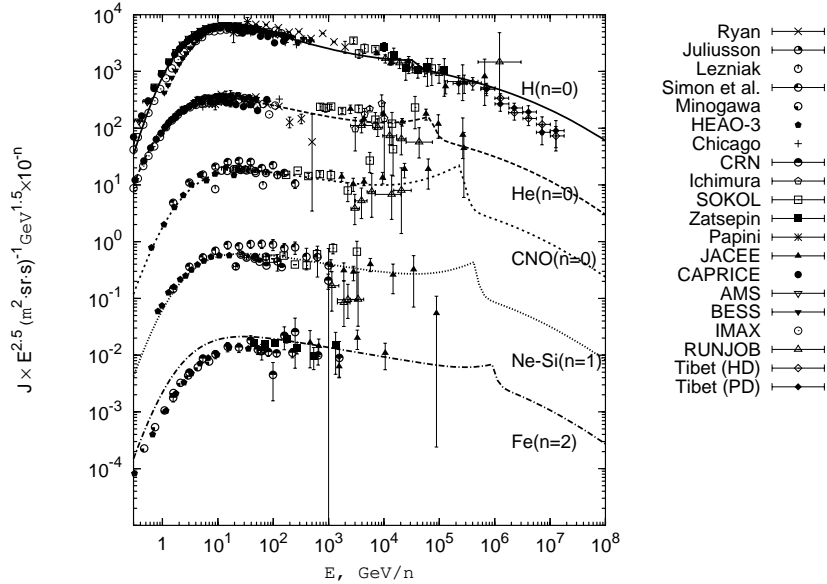}
\caption{Comparison of nuclei spectra obtained in the model with
two types of sources with experimental data (experimental data 
from~\cite{Lagutin2005cr}).} \label{allnucl} 
\end{figure*}

For punctual impulse source with power energy spectrum
$S(\vec{r},t,R)=S_0R^{-p}\delta(\vec{r})\Theta(T-t)\Theta(t)$, corresponding to
particle generation processes in astrophysical objects, the solution of
equation~(1) has the form
\begin{eqnarray}\label{nrtE}
\nonumber
N(\vec{r},t,R)=\frac{S_0R^{-p}}{D(R,\alpha,\beta)^{3/\alpha}}\int\limits_{\max[0,t-T]}^{t}\tau^{-3\beta/\alpha}\times\\
\times
\Psi_3^{(\alpha,\beta)}\Bigl({|\vec{r}|(D(R,\alpha,\beta)\tau^\beta)^{-1/\alpha}}\Bigr)d\tau,
\end{eqnarray}
where function $\Psi_3^{(\alpha,\beta)}(r)$ --- density of fractional stable
distribution~\cite{Uchaik2003ufn}.

Analysis of the obtained solution and its comparison with the available 
experimental data for energies below and above the `knee' allowed to determine 
the basic model parameters~\cite{Lagutin2005cr}. It is shown, that for 
$p\approx 2.85$, $\delta \approx 0.27$, anomalous diffusion model satisfactory 
describes the basic observable features of cosmic ray spectrum and mass 
composition in the wide energy range $10^2-10^{10}$~(see, 
e.g.,~\cite{Lagutin2003an,Lagutin2005cr}). The obtained all-particle spectrum 
for sources with $S(E)\sim E^{-2.85}$ practically coincides with `background' 
spectrum from paper~\cite{Erlyki2001ic} (Figure~\ref{ew1}). At the same time 
the calculations demonstrated that for purely power-like spectrum $S\sim 
E^{-2.85}$ the model is unable to reproduce additional breaks in the region 
$10^5-10^7$~GeV. 

\begin{figure}
\centering\includegraphics[angle=-90,width=.5\textwidth]{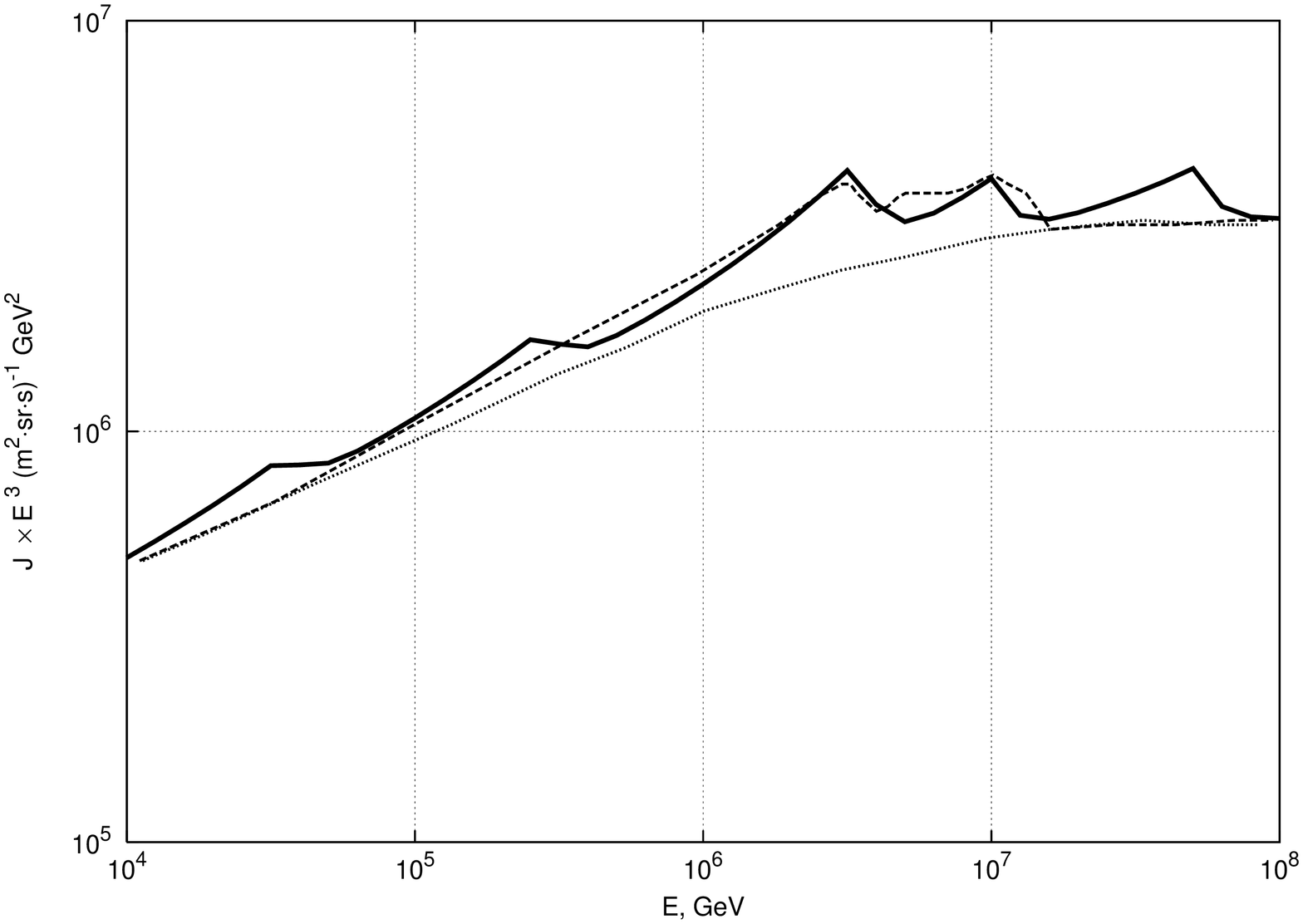}
\caption {Cosmic ray spectrum in the `knee' region. Comparison of
the data~\cite{Erlyki2001ic} with the present paper results for the model with 
two types of sources. Dotted line -- `background' spectrum~\cite{Erlyki2001ic}, 
dashed line -- spectrum~\cite{Erlyki2001ic} with account for young nearby 
supernova, solid line -- all-particle spectrum, obtained in the present paper 
within the model with two types of sources.} 
\label{ew2}
\end{figure}

\begin{figure}
\centering\includegraphics[angle=-90,width=0.5\textwidth]{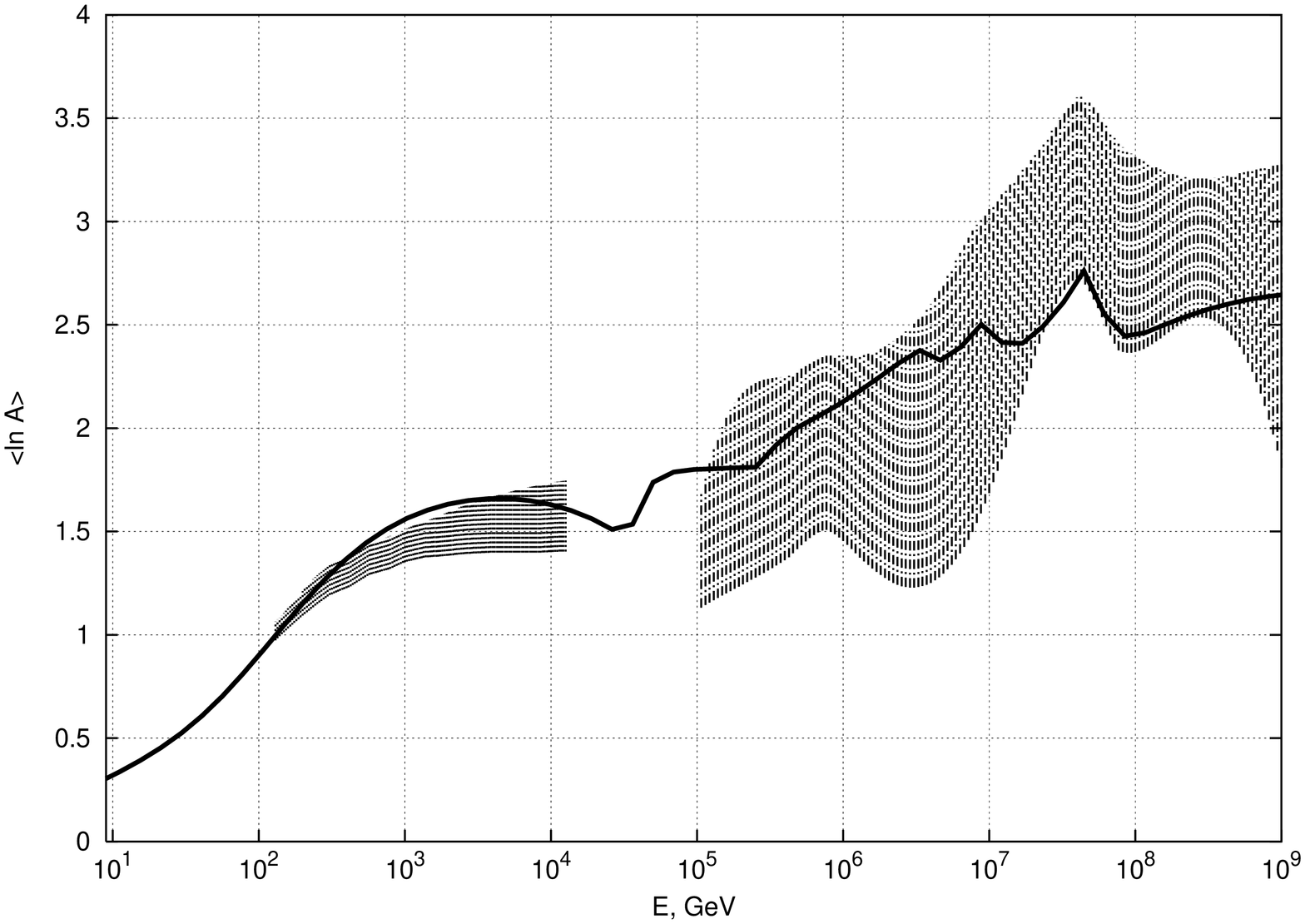}
\caption{Mean logarithmic cosmic ray mass vs.~primary particle
energy. Dashed areas --- experimental data from~\cite{Shibat1999, Hoeran2003}, 
solid line --- our calculation in the framework of the anomalous diffusion 
model with two types of sources.} 
\label{lna}
\end{figure}

\section{Results and conclusions}

To find conditions when the model, formulated in Introduction, is able to 
reproduce results of~\cite{Erlyki2001ic,Hoeran2003}, spectra of nuclei, 
all-particle spectrum and $\langle\ln A\rangle$ were calculated with the use of 
(2) and particle generation spectrum in the source of the form $S(E) = S_0 
E^{-2.85} + S_1 E^{-2}\Theta(E_{\max}- E)$. As a criterion for determination of 
particle maximum energy, accelerated by supernova type source, we used 
conclusion on the considerable increase of $\langle\ln A\rangle$ in the region 
$10^7-10^8$~GeV, obtained in~\cite{Hoeran2003}. Results of our calculations,  
given in Figures~\ref{allnucl}--\ref{lna} and in Table~1, show, that inclusion 
into system of sources with $S(E)\sim E^{-2.85}$ adopted 
in~\cite{Lagutin2005cr}, of the additional supernova-type source 
($r\approx100$~pc, $t\approx10^5$~yr), accelerating particles up to 
$E_{\max}\approx3\cdot10^4Z$~GeV with output to proton component $\sim 2\cdot 
10^{48}$~erg/source, allows to describe complicated structure of cosmic ray 
spectrum and mass composition in the vicinity of the `knee'. As mentioned 
above, the all-particle spectrum for sources with $S(E)\sim E^{-2.85}$ 
practically coincides with `background' spectrum from 
paper~\cite{Erlyki2001ic}.

\begin{table}
\caption{Mean logarithmic cosmic ray mass $\langle \ln A \rangle$
vs.~primary particle energy in the framework of the anomalous
diffusion model with two types of sources.}
\renewcommand{\tabcolsep}{6pt}
\begin{center}
\begin{tabular}{rrrr} \hline
E, GeV& $\langle \ln A \rangle$ &E, GeV& $\langle \ln A \rangle$
\\ \hline 
$10^1$        &0.32  & 6.3$\cdot10^5$ &2.05 \\ 
2.5$\cdot10^1$&0.49  & $10^6$         &2.13 \\            
6.3$\cdot10^1$&0.74  & 2.5$\cdot10^6$ & 2.33 \\
$10^2$        &0.90  & 6.3$\cdot10^6$ & 2.39 \\
2.5$\cdot10^2$&1.23  & $10^7$         & 2.53 \\
6.3$\cdot10^2$&1.48  & 1.3$\cdot10^7$ & 2.39 \\
$10^3$        &1.56  & 2.0$\cdot10^7$ & 2.45 \\
2.5$\cdot10^3$&1.65  & 2.5$\cdot10^7$ & 2.52 \\
6.3$\cdot10^3$&1.65  & 3.2$\cdot10^7$ & 2.60 \\
$10^4$        &1.63  & 4.0$\cdot10^7$ & 2.69 \\ 
2.5$\cdot10^4$&1.52  & 5.0$\cdot10^7$ & 2.79 \\
6.3$\cdot10^4$&1.78  & 6.3$\cdot10^7$ & 2.51 \\
$10^5$        &1.80  & 7.9$\cdot10^7$ & 2.44 \\ 
2.5$\cdot10^5$&1.81  & $10^8$&2.45 \\          
\hline
\end{tabular}
\end{center}
\end{table}

In conclusion, also, let us note, that verification of the
hypothesis~\cite{Erlyki2001ic} about contribution of nearby
supernova to the presently observed in the Solar system cosmic ray
spectrum may be accomplished via execution of additional
measurements (or complementary analysis), e.g., of H, He and CNO
nuclei spectra in some energy regions. As follows from our
calculations, presented in Figure~\ref{allnucl}, nearby supernova
could give rise to inhomogeneities in the spectra of these nuclei
in the regions of $3\cdot10^4$ GeV/nucleon (H), $6\cdot10^4$
GeV/nucleon (He), $3\cdot10^5$ GeV/nucleon (CNO). Indications on
the presence of such inhomogeneities in the spectra of He and CNO
in these energy regions are present in the data of JACEE
collaboration~\cite{jacee}.

\end{document}